\documentclass{elsart5}
\usepackage{graphicx}
\newlength{\figwidth} 
\usepackage{amssymb}
\begin{document}
\begin{frontmatter}
\title{
Anomalous magnetic properties near Mott transition \\
in Kagom{\'e} lattice Hubbard model}
\author[aff1]{T. Ohashi\corauthref{cor1}\thanksref{label1}}
\ead{t-ohashi@riken.jp}
\thanks[label1]{Present address: 
Condensed Matter Theory Laboratory, 
RIKEN, Wako, Saitama 351-0198, Japan}
\corauth[cor1]{Takuma Ohashi}
\author[aff1]{S.-i. Suga}
\author[aff1]{N. Kawakami}
\author[aff2]{H. Tsunetsugu}
\address[aff1]{Department of Applied Physics, Osaka University, 
Suita, Osaka 565-0871, Japan}
\address[aff2]{Yukawa Institute for Theoretical Physics, 
Kyoto University, Kyoto 606-8502, Japan}
\received{\today}

\begin{abstract}
We investigate the characteristics of the metallic phase 
near the Mott transition in the Kagom\'e lattice Hubbard model
using the cellular dynamical mean field theory. 
By calculating the specific heat and spin correlation functions, 
we demonstrate that the quasiparticles show anomalous properties 
in the metallic phase close to the Mott transition. 
We find clear evidence for the multi-band 
heavy quasiparticles in the specific heat, 
which gives rise to unusual temperature dependence 
of the spin correlation functions. 
\end{abstract}
\end{frontmatter}

\begin{figure}[bt]
\begin{center}
\includegraphics[clip,width=0.38\textwidth]{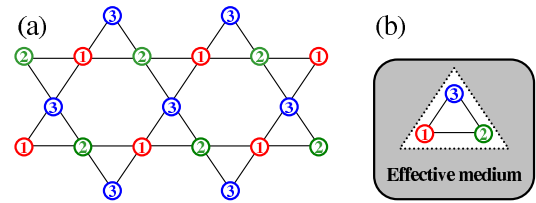}
\end{center}
\caption{
(a) Sketch of the Kagom\'e lattice and 
(b) the effective cluster model using three sites cluster CDMFT. 
}
\label{fig:kagome}
\end{figure}

Geometrical frustration has attracted much attention 
in the field of strongly correlated electron systems. 
The observation of heavy fermion behavior in 
the pyrochlore compound $\mathrm{LiV_2O_4}$
\cite{kondo97} and the discovery of superconductivity 
in the triangular lattice compound 
$\mathrm{Na_xCoO_2 \cdot yH_2O}$ \cite{takada03}, 
the $\beta$-pyrochlore osmate $\mathrm{KOs_2O_6}$ \cite{yonezawa04},
etc. have stimulated intensive studies of frustrated electron systems. 
The Kagom\'e lattice (Fig. \ref{fig:kagome}(a)) 
is another prototype of frustrated systems, 
which may be regarded as a two-dimensional analog of the pyrochlore lattice. 
It is suggested that a correlated electron system on the Kagom\'e lattice 
can be an effective model of $\mathrm{Na_xCoO_2 \cdot yH_2O}$ 
\cite{koshibae03}. 
The issue of electron correlations for 
the Kagom\'e lattice systems was addressed 
by using the FLEX approximation \cite{imai03} 
and QMC method \cite{bulut05} in the metallic regime. 
In our recent paper \cite{ohashi06}, 
we have studied electron correlations in 
the Kagom{\'e} lattice Hubbard model, 
and found the first-order Mott transition 
at the Hubbard interaction $U/W \sim 1.37$ ($W$: band width). 

In this paper, we focus on the anomalous properties 
around the Mott transition in the Kagom{\'e} lattice Hubbard model. 
We consider the standard Hubbard model with 
nearest-neighbor hopping $t$ ($>0$) on the Kagom{\'e} lattice, 
\begin{eqnarray*}
H= -t \sum_{\left \langle i,j \right \rangle ,\sigma}
c_{i\sigma }^\dag c_{j\sigma}
+ U \sum_{i} c_{i\uparrow}^\dag   c_{i\uparrow} 
             c_{i\downarrow}^\dag c_{i\downarrow}, 
\end{eqnarray*}
where $c_{i\sigma }^\dag$ ($c_{j\sigma}$) creates (annihilates) 
an electron with spin $\sigma$ at site $i$. 
We use the band width $W=6t$ as a unit of energy.
In order to treat both effects of strong correlations and 
geometrical frustration, 
we employ the cellular dynamical mean field theory 
(CDMFT) \cite{kotliar01}. 
In CDMFT, the original lattice is regarded 
as a super-lattice consisting of clusters, 
which is then mapped onto an effective cluster model 
via a standard DMFT procedure. 
Each unit cell of the Kagom\'e lattice has three sites labeled by 
$1$, $2$, and $3$, as shown in Fig. \ref{fig:kagome}(a). 
We thus end up with a three-site cluster model 
coupled to the self-consistently determined medium 
illustrated in Fig. \ref{fig:kagome}(b), 
which is solved by using QMC method \cite{hirsch86}. 

\begin{figure}[bt]
\begin{center}
\includegraphics[clip,width=\figwidth]{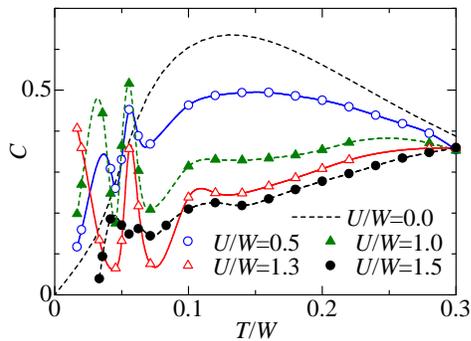}
\end{center}
\caption{
The temperature dependence of the specific heat $C$
for several values of interaction strength $U/W$. 
The Mott transition occurs at $U/W \sim 1.37$ \cite{ohashi06}. 
}
\label{fig:heat}
\end{figure}

We now investigate the characteristics around 
the Mott transition  at half filling. 
In Fig. \ref{fig:heat}, we show the temperature ($T$) dependence of 
the specific heat $C$ for several values of interaction strength. 
In the noninteracting case ($U=0$), 
$C$ has a single hump structure. 
As $U$ increases ($U/W=0.5$, $1.0$, $1.3$), 
sharp peaks are developed in the low-$T$ regime
due to low-energy spin excitations, 
while charge excitations feature the higher-$T$ hump structure. 
In particular, in the strongly correlated metallic regime 
at $U/W=1.0$ and $1.3$, we clearly find the three different structures
 at low $T$: the shoulder around $T/W \sim 0.1$, 
the peak at $T/W \sim 0.06$, and  the other sharp peak at lower $T$. 
These characteristic structures reflect
the three bands in the Kagom{\'e} lattice electron system 
which consist of one flat and two dispersive bands. 
In the small-$U$ region, the electrons near the Fermi surface 
are mainly renormalized and 
contribute to the low-energy excitations so that  
the low-$T$ peak structure is indistinct. 
In the strong-$U$ region ($U/W=1.0$, $1.3$), 
the whole three bands including the two bands away from the
Fermi surface are renormalized to participate in 
the formation of quasiparticles. 
Therefore, the three types of quasiparticles become all relevant for 
the low-energy excitations, resulting in 
the three distinct structures in $C$.
On the other hand, 
one can see the suppression of $C$ 
at low $T$ in the insulating phase ($U/W=1.5$). 
In our previous study \cite{ohashi06}, 
we found a dramatic change in spin excitation spectra at 
the Mott transition. 
The single hump structure splits into two peaks 
once the system enters the insulating phase, where 
the low-energy peak is mainly formed by the AF coupled electrons between 
adjacent sites while the high-energy one is due to
almost free spin states created in a three-site cluster. 
As $T$ decreases, the contribution of the high-energy 
spin excitations to $C$ becomes small, 
resulting in the decrease of $C$ at low $T$. 

\begin{figure}[bt]
\begin{center}
\includegraphics[clip,width=\figwidth]{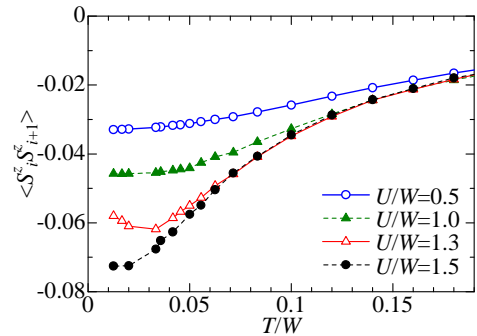}
\end{center}
\caption{
The temperature dependence of 
the nearest neighbor spin correlation function 
$\langle S^z_i S^z_{i+1} \rangle$ for several values of $U/W$. 
}
\label{fig:correlation}
\end{figure}

By investigating the short-range spin correlations, 
we further find the anomalous magnetic properties 
in the metallic phase close to the Mott transition. 
Shown in Fig. \ref{fig:correlation} is 
the $T$-dependence of the nearest-neighbor spin correlation function 
$\langle S^z_i S^z_{i+1} \rangle$ for several values of $U/W$. 
$\langle S^z_i S^z_{i+1} \rangle$ is always negative so that 
the spin correlation is antiferromagnetic (AF), 
which is a source of strong frustration 
on the frustrated Kagom{\'e} lattice. 
As $T$ decreases, the nearest-neighbor AF spin correlation 
is enhanced gradually. 
In the insulating phase ($U/W=1.5$)
the AF spin correlation becomes stronger 
with decreasing $T$, as is expected. 
On the other hand, $\langle S^z_i S^z_{i+1} \rangle$
in the metallic phase close to 
the Mott transition ($U/W=1.3$) shows 
a nonmonotonic $T$-dependence:
the AF spin correlation is once enhanced and then suppressed 
with the decrease of $T$. 
The unusual $T$-dependence results from 
the competition between the quasiparticle formation 
and the frustrated spin correlations, 
which may be characterized by two energy scales: 
the coherence temperature $T_C$ and 
$T_M$ characterizing the AF spin fluctuations. 
The AF correlation is developed around $T \sim T_M$, 
which stabilizes localized moments and causes frustration 
in accordance with the monotonic enhancement 
of spin correlations in the insulating phase.
On the other hand, 
when the system is in the metallic phase, 
electrons recover coherence in itinerant motion below $T_C$. 
Therefore, the frustration is relaxed by itinerancy of electrons
via the suppression of AF correlations at $T<T_C$. 
Thus, the nonmonotonic $T$-dependence of
$\langle S^z_i S^z_{i+1} \rangle$ clearly demonstrates that 
the heavy quasiparticles are formed 
under the influence of strong frustration. 

In summary, we have studied the Kagom\'e lattice Hubbard model 
by means of CDMFT combined with QMC. 
We have found some anomalous magnetic properties 
in the metallic phase close to the Mott transition. 
The calculation of the specific heat has demonstrated 
the formation of multi-band quasiparticles, 
which shows the unusual temperature dependence of 
the spin correlation function. 

The authors thank Y. Motome, A. Koga, and Y. Imai for valuable discussions. 
A part of numerical computations was done at the Supercomputer Center 
at ISSP, University of Tokyo and also at YITP, Kyoto university. 
This work was partly supported by Grant-in-Aid for Scientific Research
(No. 16540313) and for Scientific Research on Priority Areas
(No. 17071011 and 18043017) from MEXT.


\begin{thebibliography}{00}
\bibitem{kondo97}
S. Kondo, {\it et al.}, 
Phys. Rev. Lett. {\bf 78}, 3729 (1997). 

\bibitem{takada03}
K. Takada, {\it et al.}, 
Nature (London) {\bf 422}, 53 (2003). 

\bibitem{yonezawa04}
S. Yonezawa, {\it et al.}, 
J. Phys.: Cond. Mat. {\bf 16}, L9 (2004). 

\bibitem{koshibae03}
W. Koshibae, {\it et al.}, 
Phys. Rev. Lett. {\bf 91}, 257003 (2003). 

\bibitem{imai03}
Y. Imai, {\it et al.}, 
Phys. Rev. B {\bf 68}, 195103 (2003). 

\bibitem{bulut05}
N. Bulut, {\it et al.}, 
Phys. Rev. Lett. {\bf 95}, 037001 (2005). 

\bibitem{ohashi06}
T. Ohashi, {\it et al.}, 
Phys. Rev. Lett. {\bf 97}, 066401 (2006). 

\bibitem{kotliar01}
G. Kotliar, {\it et al}, 
Phys. Rev. Lett. {\bf 87}, 186401 (2001). 

\bibitem{hirsch86} 
J. E. Hirsch and R. M. Fye, 
Phys. Rev. Lett. {\bf 56}, 2521 (1986). 
\end{thebibliography}
\end{document}